\documentclass[aps,preprintnumbers,amsmath,amssymb,nofootinbib]{revtex4}
\usepackage{textcomp}
\usepackage{mathtools}
\usepackage{amsmath}
\usepackage{amssymb}
\usepackage{esint}
\usepackage[utf8]{inputenc}
\usepackage{hyperref}

\makeatletter

\DeclareFontEncoding{LGR}{}{}

\ProvideTextCommand{\~}{LGR}[1]{\char126#1}

\@ifundefined{textcolor}{}
{%
 \definecolor{BLACK}{gray}{0}
 \definecolor{WHITE}{gray}{1}
 \definecolor{RED}{rgb}{1,0,0}
 \definecolor{GREEN}{rgb}{0,1,0}
 \definecolor{BLUE}{rgb}{0,0,1}
 \definecolor{CYAN}{cmyk}{1,0,0,0}
 \definecolor{MAGENTA}{cmyk}{0,1,0,0}
 \definecolor{YELLOW}{cmyk}{0,0,1,0}
}

\usepackage[final]{graphics}
\usepackage{amsfonts}
\usepackage{color}
\usepackage{MnSymbol}

\def\be{\begin{equation}}
\def\ee{\end{equation}}
\def\bea{\begin{eqnarray}}
\def\eea{\end{eqnarray}}

\def\beq{\begin{equation}}
\def\eeq{\end{equation}}

\renewcommand{\v}[1]{ \ensuremath{  \underline {#1} }}

\makeatother

\begin{document}

\title {  Bose enhancement, the Liouville effective action and\\ the high multiplicity tail in p-A collisions} 

\preprint{CERN-TH-2018-101, RBRC-1281}
\author{Alex Kovner}
\affiliation{Physics Department, University of Connecticut, 2152 Hillside Road, Storrs, CT 06269, USA}
\affiliation{Theoretical Physics Department, CERN, CH-1211 Geneve 23, Switzerland}

\author{Vladimir V. Skokov}
\affiliation{RIKEN/BNL, Brookhaven National Laboratory, Upton, NY 11973, USA}

\date{\today}
\begin{abstract}
In the framework of dense-dilute CGC approach we study fluctuations in the
multiplicity of produced particles in p-A collisions. We show that the leading
effect that drives the  fluctuations is the Bose enhancement of gluons in the
proton wave function. We explicitly calculate the moment generating function
that resums the effects of Bose enhancement. We show that it can be understood
in terms of the Liouville effective action for the composite field which is
identified with the fluctuating density, or saturation momentum of the proton.
The resulting probability distribution turns out to be very close to the
$\gamma$-distribution. We also calculate the first correction to this
distribution which is due to pairwise Hanbury Brown-Twiss correlations of
produced gluons.
\end{abstract}
\date{\today}

\maketitle
\section{Introduction}
Study of correlations in p-A and p-p collisions have been a very active area in
the last years due to the observation of the ridge correlations  at LHC. Since
the ridge signal is much more pronounced in high multiplicity events, it is
very important to understand the origin of the multiplicity fluctuation and
especially the high multiplicity tail of the distribution, see
Ref.~\cite{Khachatryan:2010nk,Aad:2010ac,Aaij:2014pza,ALICE:2017pcy} for the
most recent experimental studies~\footnote{See also Refs.~\cite{Abelev:2012rz,Adam:2016mkz,Weber:2017hhm} and \cite{Ma:2018bax}
for the experimental and theoretical studies of heavy flavour production in high multiplicity collisions of small systems.}.

In the present paper we address this question in the framework of the Color
Glass Condensate (CGC) approach.
We calculate the multiplicity momentum generating function, using
McLerran-Venugopalan (MV) model for the projectile and assuming that the target
is very dense. The latter assumption allows us to employ factorizable form for
the averages of Wilson lines, as explained in Ref.~\cite{Kovner:2017vro}. We
show that the main contribution to multiplicity fluctuations arises form the
Bose enhancement (BE) of gluons in the projectile wave function (see
Ref.~\cite{Altinoluk:2015uaa} providing the interpretation of the correlations
in the projectile in terms of BE; a related effect of Pauli blocking for quarks
is discussed
in Ref.~\cite{Altinoluk:2016vax}).
This effect produces fluctuations which are not suppressed by the factor of the
area of the projectile.
We are able to resum the BE contributions exactly in the multiplicity
generating function. The resulting distribution turns out to be
$\gamma$-distribution, which for large moments practically coincides with the
negative binomial distribution.

The other important effect in correlated gluon production  is the Hanbury
Brown-Twiss (HBT) effect. As discussed at length in recent literature
\cite{Kovner:2017ssr}, it is the leading cause for the angular correlations of
produced gluons in the CGC approach. Its contribution to the total multiplicity
on the other hand is suppressed relative to that of BE, as the correlated peak
that it produces is very narrow. Nevertheless we identify the contributions to
the multiplicity generating function  due to the HBT within our calculational
framework and calculate corrections  induced by it.

We note that a calculation along similar lines has been undertaken some years
ago in Ref.~\cite{Gelis:2009wh}, see also
Refs.~\cite{Dumitru:2012yr,Schenke:2012fw,Schenke:2013dpa} for numerical
calculations and comparison to experimental data and Ref.~\cite{Dumitru:2012tw}
for
higher order correction to an MV model.
There are however significant differences between our approach and that of
Ref.~\cite{Gelis:2009wh}. In particular Ref.~\cite{Gelis:2009wh} treated both
the projectile and the target as dilute. It turns out that the large density
effects of the target suppress half of the contributions considered as leading
in Ref.~\cite{Gelis:2009wh}. Additionally the HBT contributions have not been
included in the analysis of Ref.~\cite{Gelis:2009wh}. The resulting
multiplicity distribution we obtain is somewhat different from that in
Ref.~\cite{Gelis:2009wh}.

The structure of this paper is the following. In Section II we lay out the general framework of the calculation and perform the averaging over the projectile wave function using the MV model.

 In Section III we make a detour and consider specifically the second moment of the multiplicity distribution, a.k.a. inclusive two gluon cross section. We show explicitly that BE leads to the largest contribution to this moment and that this contribution is not suppressed by a power of area relative to the square of the single inclusive cross section. The argument is very similar to that in Ref.~\cite{Gelis:2009wh} except for the differences alluded to above due to the dense nature of the target. Similar observation was made also in Ref.~\cite{Kovchegov:2013ewa}.

 In Section IV we calculate in closed form the momentum generating function which resums all BE contributions. We also relate it to the constrained effective potential approach proposed in Ref.~\cite{Dumitru:2017cwt} and show that our result for the momentum generating function  is identical to the Liouville theory for the ``composite field'' discussed in Ref.~\cite{Dumitru:2017cwt}.

 In Section V we consider the HBT corrections to the distribution and provide a closed expression that resums the leading correction.

 Finally Section VI contains a short discussion of our results.

\section{The generating function}
Our calculations will be performed
within the dense-dilute CGC framework.  In this approach the number of produced gluons for a given configuration of the projectile (proton) and
a target (nucleus) is given by~\cite{Kovchegov:1998bi,Dumitru:2001ux,Blaizot:2004wu}
\begin{equation}
\left.\frac{dN}{d^{2}kdy}\right|_{\rho_{\rm p},\rho_{\rm t}}=\frac{2g^{2}}{(2\pi)^{3}}\int\frac{d^{2}q}{(2\pi)^{2}}\frac{d^{2}q'}{(2\pi)^{2}}\Gamma(\v{k},\v{q},\v{q}')\rho_{\rm p}^{a}(-\v{q}')\left[U^{\dagger}(\v{k}-\v{q}')U(\v{k}-\v{q})\right]_{ab}\rho_{\rm p}^{b}(\v{q}),
\end{equation}
where the square of Lipatov vertex is
\begin{equation}
\Gamma(\v{k},\v{q},\v{q}')=\left(\frac{\v{q}}{q^{2}}-\frac{\v{k}}{k^{2}}\right)
\cdot \left(\frac{\v{q}'}{q'^{2}}-\frac{\v{k}}{k^{2}}\right)\,.
\end{equation}
Here $\rho_{\rm p}$ is a given configuration of the color charged density in the projectile, and $U$ is the eikonal scattering matrix -- the adjoint Wilson line -- for scattering of a single gluon on the target.
The target Wilson lines depend on the target color sources, $\rho_{\rm t}$; we suppress this in our notation.

The single inclusive and double inclusive production in this approach
are given by
\begin{equation}
	\left.\frac{dN}{d^{2}kdy}=\left\langle \left\langle  \frac{dN}{d^{2}kdy}\right|_{\rho_{\rm p},\rho_{\rm t}}
	\right \rangle_{\rm p}
	\right\rangle_{\rm t}
\end{equation}
and
\begin{equation}
\frac{d^{2}N}{d^{2}k_{1}dy_{1}d^{2}k_{2}dy_{2}}=
\left\langle \left \langle
\left. \frac{dN}{d^{2}k_{1}dy_{1}}\right|_{\rho_{\rm p},\rho_{\rm t}}
\left.\frac{dN}{d^{2}k_{2}dy_{2}} \right|_{\rho_{\rm p},\rho_{\rm t}}
 \right\rangle_{\rm p}
 \right\rangle_{\rm t} \, ,
\end{equation}
where the averaging is performed over the projectile and target color charge configurations:
\begin{equation}
	\left\langle O(\rho_{\rm p})
 	\right\rangle_{\rm p}  =
	\frac{1}{Z_{\rm p}} \int {\cal D} \rho_{\rm p}\  W_{\rm p}(\rho_{\rm p})\  O(\rho_{\rm p})
\end{equation}
and
\begin{equation}
	\left\langle O(\rho_{\rm t})
 	\right\rangle_{\rm t}    =
	\frac{1}{Z_{\rm t}} \int {\cal D} \rho_{\rm t}\  W_{\rm t}(\rho_{\rm t})\  O(\rho_{\rm t})\,.
\end{equation}
The normalization factors, $Z_{\rm p,t}$, are fixed so that
\begin{equation}
	\left\langle 1
 	\right\rangle_{\rm p}
	=	\left\langle 1
	\right\rangle_{\rm t} =1\,.
\end{equation}
In general $m$-particle production is
\begin{equation}
\frac{d^{m}N}{d^{2}k_{1}dy_{1}d^{2}k_{2}dy_{2}  \dots d^{2}k_{m}dy_{m}  }=
\left\langle  \left \langle
\left. \frac{dN}{d^{2}k_{1}dy_{1}}\right|_{\rho_{\rm p},\rho_{\rm t}}
\left. \frac{dN}{d^{2}k_{2}dy_{2}}\right|_{\rho_{\rm p},\rho_{\rm t}}
\ldots
\left.\frac{dN}{d^{2}k_{m}dy_{m}} \right|_{\rho_{\rm p},\rho_{\rm t}}
 \right\rangle_{\rm p}
 \right\rangle_{\rm t} \,.
\end{equation}

Instead of computing the moments of inclusive particle number fluctuations we
 evaluate the moment  generating function (see e.g. Ref.~\cite{Kovner:2006wr})
 \begin{equation}
	 G(t)
	 =
\left\langle  \left \langle
 \exp\left[
 t
 \int_{k_{\rm min}} d^2 k
\left.\frac{dN}{d^{2}kdy}\right|_{\rho_{\rm p},\rho_{\rm t}}
 \right]
\right\rangle_{\rm p}
\right
\rangle_{\rm t}\,,
 \end{equation}
where we introduced an arbitrary $k_{\rm min} \gg \Lambda_{\rm QCD}$.
The moments of the distribution are obviously obtained from $G(t)$ by differentiating with respect to $t$ at $t=0$.

To calculate the generating function we have to specify the distribution of the sources in the projectile and in the target.
For the projectile we will use the simple Gaussian McLerran-Venugopalan (MV) model specified by
\begin{equation}
	\langle \rho_{\rm p}^a(\v{p}) \rho_{\rm p}^b(\v{k})  \rangle_{\rm p}
	= (2\pi)^2  \mu_{\rm p} ^2(p)  \delta(\v{p}+\v{k}) \delta^{ab}\,
	\label{Eq:MVp}
\end{equation}
which corresponds to the weight functional
\begin{equation}
	W_{\rm p}(\rho_{\rm p}) = \exp\left( - \int \frac{d^2 q}{(2\pi)^2}
	\rho_{\rm p}^a(-\v{q})
	\frac{1}{2\mu_{\rm p}^2(q)}
	\rho_{\rm p}^a(\v{q})
	\right) \,.
\end{equation}
Note that the structure of the $\rho_{\rm p}$ correlator means translational invariance of the projectile wave function in the transverse plane. This assumption is only reasonable if we concentrate on momenta of produced particles larger than the inverse radius of the projectile. Thus in the following we will always assume $k_{\rm min}>1/R$.

The averaging over the Wilson lines of the target will be performed in the approximation articulated recently in Ref.~\cite{Kovner:2017ssr}. Any product of Wilson lines is factored into pairs with the basic Wick contraction
\begin{equation}
	\left\langle U_{ab}(\v{p})U_{cd}(\v{q}) \right\rangle_{\rm t} = \frac{(2\pi)^2}{N_{c}^{2}-1}\delta_{ac}\delta_{bd}\delta(\v{p}+\v{q}) D(\v{p})\,.
\end{equation}
Here the adjoint dipole amplitude is defined as
\begin{equation}
	D(p) = \frac{1}{N_c^2-1}\int d^2x e^{ix\cdot p}\langle{\rm tr} \ \left[ U^\dagger(x)  U(0) \right] \rangle_{\rm t}\,.
\end{equation}
As explained in Ref.~\cite{Kovner:2017ssr} this approximation is appropriate for the dense 
 regime. It collects all terms in the $n$-particle cross section which have the leading dependence on the area of the projectile.  The approximation only includes terms which contain ``small size''  color singlets in the projectile propagating through the target. Any non singlet state that in the transverse plane is removed by more than $1/Q_s$ from other propagating partons must have a vanishing $S$-matrix on the dense (black disk) target. On the other hand if the singlet state contains more than two partons, one looses a power of the area when integrating over the coordinates of the partons. Thus the leading contribution in the black disk limit is the one where only dipole contribution to the $S$-matrix should be accounted for.
The same approximation for the quadrupole amplitude has been used previously in Ref.~\cite{Kovchegov:2013ewa} where its consistency with explicit modeling of the Wilson line correlators via MV model has been verified.

Note that this averaging procedure for the target is formally (disregarding subtleties related to the definition of the Haar measure)  equivalent to the following form of the weight functional:
\begin{equation}
W_t[U]=\exp\left\{-\frac{1}{2}\int \frac{d^2q}{(2\pi)^2}\frac{1}{D(q)}{\rm tr}[U^\dagger(q)U(q)]\right\}\,.
\end{equation}

\subsection{Projectile averaging}
We now consider the calculation of the  generating function
\begin{equation}
	G(t)=\int D\rho_{\rm p} DU \exp\left[ - \int \frac{d^2 q}{(2\pi)^2} \left(
		\rho_{\rm p}^a(-\v{q})
	\frac{1}{2\mu_{\rm p}^2(q)}
	\rho_{\rm p}  ^a(\v{q})+\frac{1}{2D(q)}{\rm tr}[U^\dagger(q)U(-q)]\right)+t
 \int_{k_{\rm min}} d^2 k
\left.\frac{dN}{d^{2}kdy}\right|_{\rho_{\rm p},\rho_{\rm t}}
 \right] \,.
\end{equation}
We notice immediately that, since $\frac{dN}{d^{2}kdy}$ is quadratic in both $\rho_{\rm p}$ and $U$, and both the corresponding weight functionals are Gaussian as well, we can integrate over one of these ``fields'' exactly. We choose to integrate first over $\rho_{\rm p}$. The result of this integration is
\begin{equation}\label{traced}
G(t)=\int DU\exp\left[ - \int \frac{d^2 q}{(2\pi)^2} \frac{1}{2D(q)}{\rm tr}[U^\dagger(q)U(-q)]-\frac{1}{2}{\rm tr}\ln\Big[1-tM\Big]\right]
\end{equation}
where the operator $M$ is defined by its matrix elements
\begin{equation}
M_{ab}(q',q)=\frac{4g^2}{(2\pi)^3}\mu^2(q)\int_{k_{\rm min}}   \frac{d^2k}{(2\pi)^2}\Gamma(k,q,q')\left[U^{\dagger}(\v{k}-\v{q}')U(\v{k}-\v{q})\right]_{ab}\,.
\end{equation}

At this point we need to make some approximations in order to perform the remaining functional integral over $U$. In the next section we will consider more closely the two gluon production cross section, which will help us understand the systematics of the leading contributions to the multiplicity fluctuations, and to devise the appropriate approximation that sums these leading contributions.





\section{Double inclusive production: dissecting different contributions}
To get some insight of which effects contribute the most to the multiplicity fluctuations we make a short detour and consider the double inclusive gluon production.
\begin{equation}
\frac{d^{2}N}{d^{2}k_{1}dy_{1}d^{2}k_{2}dy_{2}}=
\left\langle \left \langle
\left. \frac{dN}{d^{2}k_{1}dy_{1}}\right|_{\rho_{\rm p},\rho_{\rm t}}
\left.\frac{dN}{d^{2}k_{2}dy_{2}} \right|_{\rho_{\rm p},\rho_{\rm t}}
 \right\rangle_{\rm p}
 \right\rangle_{\rm t} \, .
\end{equation}
\subsection{Dipole contribution}
First consider the average with respect to projectile  inside
each  factor $\left.\frac{dN}{d^{2}kdy}\right|_{\rho_{\rm p},\rho_{\rm t}}$,
that is
\begin{equation}
	\left[ \frac{d^{2}N}{d^{2}k_{1}dy_{1}d^{2}k_{2}dy_{2}} \right]_{\rm Dipole} =
\left\langle \left \langle
\left. \frac{dN}{d^{2}k_{1}dy_{1}}\right|_{\rho_{\rm p},\rho_{\rm t}}
 \right\rangle_{\rm p}
\left\langle  \left.\frac{dN}{d^{2}k_{2}dy_{2}} \right|_{\rho_{\rm p},\rho_{\rm t}}
 \right\rangle_{\rm p}
 \right\rangle_{\rm t} \, .
\end{equation}
The projectile averaging gives
\begin{equation}
\left \langle
\left.\frac{dN}{d^{2}k_{1}dy_{1}}\right|_{\rho_{p},\rho_{\rm t}}
\right\rangle_{\rm p}
=\frac{2g^{2}}{(2\pi)^{3}}\int
\frac{d^{2}q}{(2\pi)^{2}}
\mu_{\rm p}^{2}(\v{q})
\Gamma(\v{k}_{1},\v{q},\v{q})\ {\rm tr}
\left[U^{\dagger}(\v{k}_{1}-\v{q})U(\v{k}_{1}-\v{q})\right]\,.
\end{equation}
The subsequent target averages give two distinct contributions.The first one involves target contractions inside each single inclusive factor and is disconnected. It reproduces the square of the single gluon production probability, and is of no interest to us.
The connected contribution is
\begin{eqnarray}\label{dd}
&&	\left \langle {\rm tr}
		    \left[U^{\dagger}(\v{k}_{1}-\v{q}')U(\v{k}_{1}-\v{q})\right]
			{\rm tr}\left[U^{\dagger}(\v{k}_{2}-\v{p}')U(\v{k}_{2}-\v{p})\right] \right\rangle^{\rm conn.}_{\rm t} =\\&&
	2 S_\perp \delta( \v{k}_{1} + \v{k}_{2}  -\v{q}  -\v{p}  )
	D^2(\v{k}_{1}-\v{q})\,.
	\notag
\end{eqnarray}
The result is of order $N_c^0$. As we will see in the following the contractions of $\rho$ that break the factors of
 $\left.\frac{dN}{d^{2}kdy}\right|_{\rho_{\rm p},\rho_{\rm t}}$ are of order $N_c^2$ and thus are more important.
 The contribution from Eq.~(\ref{dd}) can therefore be neglected at large $N_c$, and we will not try to include it and analogous contributions for higher moments  in the generating function.
  The physics of this contribution was discussed in Ref.~\cite{Altinoluk:2018hcu}. There it was shown that it corresponds to the Bose enhancement of gluons in the {\em target} wave function. Note that in the framework of the dilute target expansion utilized in Ref.~\cite{Gelis:2009wh} this contribution is leading order in $N_c$. The different $N_c$ counting for the same quantity in the dense and dilute limits is not uncommon in saturation approaches.

  \subsection{The quadrupole contributions}
  We now concentrate on the other two contractions of the projectile color charges.
 There are two such contractions, and they both lead to a single trace ``quadrupole'' contribution the production probability:
 \begin{align}
&\left\langle
\overbrace{\rho_{\rm p}  ^{a}(-\v{q}')[U^{\dagger}(\v{k}_1-\v{q}')U(\v{k}_1-\v{q})]_{ab}\rho _{\rm p}  ^{b}(\v{q})}^{dN/d^2k_1dy_1}
\overbrace{\rho _{\rm p} ^{c}(-\v{p}')[U^{\dagger}(\v{k}_2-\v{p}')U(\v{k}_2-\v{p})]_{cd}\rho _{\rm p} ^{d}(\v{p})}^{dN/d^2k_2dy_2}
\right\rangle_{\rm p}
\\ \notag & -
\left\langle
\overbrace{\rho _{\rm p} ^{a}(-\v{q}')[U^{\dagger}(\v{k}_1-\v{q}')U(\v{k}_1-\v{q})]_{ab}\rho _{\rm p} ^{b}(\v{q})}^{dN/d^2k_1dy_1}
\right\rangle_{\rm p}
\left\langle
\overbrace{\rho _{\rm p} ^{c}(-\v{p}')[U^{\dagger}(\v{k}_2-\v{p}')U(\v{k}_2-\v{p})]_{cd}\rho _{\rm p} ^{d}(\v{p})}^{dN/d^2k_2dy_2}
\right\rangle_{\rm p}
\\ \notag  & =
(2\pi)^2 \mu_{\rm p}^2(\v{p})
\delta(\v{p} - \v{q}')
(2\pi)^2 \mu_{\rm p}^2(\v{q})
\delta(\v{q} - \v{p}')
{\rm tr}
[U^{\dagger}(\v{k}_1-\v{p})U(\v{k}_1-\v{q})
U^{\dagger}(\v{k}_2-\v{q})U(\v{k}_2-\v{p})]
\\ \notag &+
(2\pi)^2 \mu_{\rm p}^2(-\v{p}')
\delta(-\v{p}' - \v{q}')
(2\pi)^2 \mu_{\rm p}^2(\v{q})
\delta(\v{q} + \v{p})
{\rm tr}
[U^{\dagger}(\v{k}_1-\v{q}')U(\v{k}_1-\v{q})
U^{\dagger}(-\v{q} -  \v{k}_2)U(-\v{q}' -  \v{k}_2)]\,.
 \end{align}
Each term has two contractions  with respect to $U$ of order $N_c^2$. We organize them according to their physical meaning~\cite{Altinoluk:2018hcu}.

\subsubsection{The HBT term}
The following contraction leads to the HBT contribution (cyclic property of trace was used)
\begin{align}
&
(2\pi)^4 \mu_{\rm p}^2(\v{p})
\mu_{\rm p}^2(\v{q})
\delta(\v{p} - \v{q}')
\delta(\v{q} - \v{p}')
{\rm tr}
[ \langle {U(\v{k}_2-\v{p}) U^{\dagger}(\v{k}_1-\v{q}')} \rangle_{\rm t}
\langle {U(\v{k}_1-\v{q}) U^{\dagger}(\v{k}_2-\v{p}')} \rangle_{\rm t} ]
\\ \notag &+
(2\pi)^4 \mu_{\rm p}^2(-\v{p}')
\mu_{\rm p}^2(\v{q})
\delta(-\v{p}' - \v{q}')
\delta(\v{q} + \v{p})
{\rm tr}
[ \langle{U(\v{p}' -  \v{k}_2) U^{\dagger}(\v{k}_1-\v{q}')}\rangle_{\rm t}
\langle{U(\v{k}_1-\v{q}) U^{\dagger}(\v{p} -  \v{k}_2)}\rangle_{\rm t} ]
\\\notag & =
(N_c^2-1)
(2\pi)^8
S_\perp
\Bigg[
	\mu_{\rm p}^2(\v{p})
	\mu_{\rm p}^2(\v{q})
D(\v{k}_1-\v{p})
D(\v{k}_1-\v{q})
\delta(\v{k}_2-\v{k}_1)
+\mu_{\rm p}^2(-\v{p}')
\mu_{\rm p}^2(\v{q})
D(\v{p}'+\v{k}_1)
D(\v{k}_1-\v{q})
\delta(\v{k}_1+\v{k}_2)
\Bigg]\,.
\end{align}
Substituting into double inclusive production we get
\begin{align}\label{hbt}
&
\left[ \frac{d^{2}N}
{dy_{1}dy_{2}}
\right]_{\rm HBT}
=
2 (N_c^2-1) S_\perp
\left(\frac{2g^{2}}{(2\pi)^{3}}\right)^2
\\ \notag & \times
\int_{k_{\rm min}} d^2k
\int d^{2}q  d^{2}p \
\Gamma(\v{k},\v{q},\v{p})
\Gamma(\v{k},\v{p},\v{q})
\mu_{\rm p}^2(\v{q})
\mu_{\rm p}^2(\v{p})
D(\v{k}-\v{q})
D(\v{k}-\v{p})\,.
\end{align}

\subsubsection{Bose  enhancement in the projectile}
The remaining contraction reflects Bose enhancement of gluons in the projectile wave function:
\begin{align}
&
(2\pi)^4 \mu_{\rm p}^2(\v{p})
\mu_{\rm p}^2(\v{q})
\delta(\v{p} - \v{q}')
\delta(\v{q} - \v{p}')
{\rm tr}
[
	\langle {U^{\dagger}(\v{k}_1-\v{q}')U(\v{k}_1-\v{q})} \rangle_{\rm t}
\langle {U^{\dagger}(\v{k}_2-\v{p}')U(\v{k}_2-\v{p})} \rangle_{\rm t} ]
\\ \notag &+
(2\pi)^4 \mu_{\rm p}^2(-\v{p}')
\mu_{\rm p}^2(\v{q})
\delta(-\v{p}' - \v{q}')
\delta(\v{q} + \v{p})
{\rm tr}
[
	\langle {U^{\dagger}(\v{k}_1-\v{q}')U(\v{k}_1-\v{q})} \rangle_{\rm t}
\langle {U^{\dagger}(\v{p} -  \v{k}_2)U(\v{p}' -  \v{k}_2)} \rangle_{\rm t}  ]
\\ \notag & =
 (2\pi)^8 (N_c^2-1) S_\perp
\Bigg[
	\mu_{\rm p}^2(\v{p})
	\mu_{\rm p}^2(\v{q})
D(\v{k}_1-\v{q})
D(\v{k}_2-\v{q})
\delta(\v{q}' - \v{q})
\delta(\v{p}' - \v{p})
\delta(\v{p} - \v{q})
\\\notag & +
\mu_{\rm p}^2(-\v{p})
\mu_{\rm p}^2(\v{q})
D(\v{k}_1-\v{q})
D(\v{p}-\v{k}_2)
\delta(\v{q}' - \v{q})
\delta(\v{p}' - \v{p})
\delta(\v{p} + \v{q})
\Bigg]\,.
 \end{align}
Again substituting into double inclusive production
\begin{align}\label{be}
&
\left[ \frac{d^{2}N}
{dy_{1}dy_{2}}
\right]_{\rm BE}
=
2 (N_c^2-1) S_\perp
\\ \notag & \times
\int d^{2}q\
|\mu_{\rm p}^2(\v{q})|^2
\left|
\frac{2g^{2}}{(2\pi)^{3}}
\int_{k_{\rm min} } d^2k
\Gamma(\v{k},\v{q},\v{q})
D(\v{q}-\v{k}) \right|^2 \,.
\end{align}

\subsection{Conclusions on double inclusive}
Comparing the HBT term Eq.~(\ref{hbt}) and the BE term Eq.~(\ref{be}), we see immediately that Eq.~(\ref{be}) gives the leading contribution to the multiplicity fluctuations.

The dominant contribution  to the integral over $q$ in Eq.~(\ref{be})  comes from small $q$ due to the IR divergence of the Lipatov vertex
\begin{equation}
	\Gamma(\v{k},\v{q},\v{q}) =
	\frac{ (\v{k} - \v{q})^2  }{k^2 q^2}\, .
	%
\end{equation}
Thus  approximately we have
\begin{align}
&
\left[ \frac{d^{2}N}
{dy_{1}dy_{2}}
\right]_{\rm BE}
\approx
2 (N_c^2-1) S_\perp
\int d^{2}q\
\frac{|\mu_{\rm p}^2(\v{q})|^2}{q^4}
\left|
\frac{2g^{2}}{(2\pi)^{3}}
\int_{k_{\rm min} } d^2k
D(\v{k}) \right|^2 \,.
\end{align}
For the MV model,  $\mu_{\rm p}^2 = {\rm const}$ we get a strong IR divergence of
the integral $ \int d^{2}q\
\frac{\mu_{\rm p}^4}{q^4}$.
This divergence is regularized by the momentum scale inversely proportional to
the projectile size $\Lambda  = \frac{1}{R_{\rm p}}$, that is modulo constant factors
\begin{equation}
	 \int d^{2}q\
	 \frac{\mu_{\rm p}^4}{q^4} \propto \mu_{\rm p}^4 S_\perp \,,
\end{equation}
where $S_\perp$ is the area of the projectile.
Thus the area dependence of the BE contribution to the double inclusive cross section is the same as that of the single inclusive cross section squared. This feature has been noted and discussed earlier in Ref.~\cite{Gelis:2009wh} and Ref.~\cite{Kovchegov:2013ewa}, see also Ref.~\cite{Kovchegov:2018jun} where the
first saturation correction in the projectile to the double inclusive production was derived.

The behavior of the HBT contribution is different. There is no quadratic  IR divergence for the integration over $q$ or $p$ in Eq.~(\ref{hbt}). As long as $k_{\rm min} \gg \Lambda$; no extra factor of area arises, and thus the HBT term is subleading. We note that if we include the soft scales in the $k$ integral, that is  take $k_{\rm min}\sim \Lambda$ the situation changes and the HBT effect becomes as important as the Bose enhancement. We will not consider this situation in the present work.

Although we have concentrated on the double inclusive cross section, it is easy to see that the analysis generalizes to the higher gluon production as well~\cite{Gelis:2009wh}. For production of $n$ gluons, the BE term which has all contractions of the Wilson loops within the same single inclusive operator yields the highest power of area $S_\perp^{n-1}$. Thus in the leading approximation we will only keep these terms. We will discuss the first corrections to this approximation later on.

\section{Moment generating function - the Bose enhancement contribution}



\subsection{Leading contribution from BE}
We now return to Eq.~(\ref{traced}). We do not know how to perform the integration over $U$ in full generality. However in view of the conclusion of the previous section, we will first only keep the leading BE contributions. It is quite clear how to do that. The leading BE contribution corresponds to contracting the two Wilson lines within the same single inclusive gluon operator. In the context of Eq.~(\ref{traced}) this corresponds simply to contractions between the two Wilson lines inside the trace of the logarithm in the exponent:
\begin{eqnarray}\label{glo}
G_{\rm LO}(t)&=&\exp\left[ -\frac{1}{2}{\rm tr}\ln\Big[1-t\langle M\rangle_t\Big]\right]\nonumber\\
&=&\exp\left[-
	 \frac{1}{2}(N_c^2-1) S_\perp
	 \int_\Lambda^{k_{\rm min}}  \frac{d^{2}q}{(2\pi)^2}
\ln \left(
1 - t\  \frac{\mu_{\rm p}^2(q) {\mathfrak D}}{q^2}
\right)
 \right]\,, 
\end{eqnarray}
where we have defined the integrated dipole operator:
\begin{equation}\label{funnyd}
	{\mathfrak D} =
\frac{4g^{2}}{(2\pi)^{3}}
	\int_{k_{\rm min}} d^2 k D(\v{k}) \,.
\end{equation}

In Eq.~(\ref{glo}) we have approximated the square of the Lipatov vertex by its leading term at low $q$,\ $\Gamma(k,q,q)\approx1/q^2$,  and have restricted the integration over $q$ by $k_{\rm min}$ from above. These restrictions can be in principle lifted, but as long as $k_{\rm min}\gg \Lambda$ Eq.~(\ref{glo}) faithfully represents all leading area terms in any moment of the probability distribution.




 For the MV model $\mu_{\rm p}^2 = {\rm const}$,  the integral can be performed analytically, as follows
 \begin{align}\label{gd}
&	 G_{\rm LO} (t)
	 =
 \exp\left[
-
\frac{1}{8 \pi}
\ (N_c^2-1)
S_\perp \int_{\Lambda^2}^{k^2_{\rm min}}  {d q^2}
\ln \left(
1 - t\  \frac{\mu_{\rm p}^2 {\mathfrak D}}{q^2}
\right)
 \right]
 \\ \notag &
 = \exp\left[ -
	 \frac{1}{8 \pi} (N_c^2-1) S_\perp\left(
k_{\rm min}^2 \ln \left\{ 1 -   t\  \frac{\mu^2 {\mathfrak D}}{k_{\rm min}^2}
\right\}
- \Lambda^2 \ln \left\{ 1 -   t\  \frac{\mu^2 {\mathfrak D}}{\Lambda^2}
  \right\}
- t \mu^2 {\mathfrak D}  \ln \frac{  k_{\rm min}^2  -   t\ \mu^2 {\mathfrak D}}
{
 \Lambda^2  - t\  \mu^2 {\mathfrak D}
}
	 \right)\right]
 \\ \notag &=
 \exp\left[
	 \frac{1}{8 \pi} (N_c^2-1) S_\perp
	 \mu_{\rm p}^2 {\mathfrak D}
	 \left\{
 t \ln \frac{k^2_{\rm min}}{\Lambda^2}
+
	 \sum_{n=2}^\infty\
\frac{t^n}{n (n-1)} \
	\left(
	\left(\frac{ \mu_{\rm p}^2 {\mathfrak D}  }{\Lambda^{2}}\right)^{n-1}
	- \left(\frac{ \mu_{\rm p}^2 {\mathfrak D}  }{k_{\rm min}^{2}}\right)^{n-1}
\right)
\right\}
 \right] \,.
 \end{align}
{\color{red} ALEX:} Here when computing the momentum integrals we assumed that IR divergences are regulated on the momentum scale of order $\Lambda$. Although this is reasonable in our dilute-dense approach, 
one may expect that higher order density correction may lead to modification of this IR momentum scale to the projectile saturation momentum $Q^p_s$, if $Q^p_s> \Lambda$.

The generating function for the cumulants  is
\begin{align}
	\label{agd}
	\ln  G_{\rm LO} (t) &=
	 \frac{1}{8 \pi} (N_c^2-1) S_\perp
	 \mu_{\rm p}^2 {\mathfrak D}\
	 \left[
	 \ln \frac{k^2_{\rm min}}{\Lambda^2}
	\  t
	+
      \sum_{n=2}^\infty \frac{ 1}{n(n-1)}
	\left\{
		\left(\frac{ \mu_{\rm p}^2 {\mathfrak D}  }{\Lambda^{2}}\right)^{n-1}
		- \left(\frac{ \mu_{\rm p}^2 {\mathfrak D}  }{k_{\rm min}^{2}}\right)^{n-1}
\right\}
	 t^n
 \right]\\\notag
&\approx
\frac{1}{8 \pi} (N_c^2-1) S_\perp
\mu_{\rm p}^2 {\mathfrak D}\
	 \left[
	 \ln \frac{k^2_{\rm min}}{\Lambda^2}
	\  t
	+
      \sum_{n=2}^\infty \frac{ 1}{n(n-1)}
	  \left(\frac{ \mu_{\rm p}^2 {\mathfrak D}  }{\Lambda^{2}}\right)^{n-1}
	 t^n
 \right]\,.
\end{align}

This is our result for the multiplicity moment generating function which resums all contributions due to the Bose enhancement of gluons in the projectile wave function. We now note some of its properties.

\subsection{Some properties of the distribution}
First of all, the mean number of particles being the first moment of the distribution, is given by the first derivative of   $G(t)$ at $t=0$ and is proportional to the projectile $\mu_{\rm p}^2$ and
the logarithm of ratio of the momentum scales as expected
\begin{equation}\label{fm}
\kappa_1= \frac{1}{8 \pi} (N_c^2-1) S_\perp
	 \mu_{\rm p}^2 {\mathfrak D}
	 \ln \frac{k^2_{\rm min}}{\Lambda^2}\,.
\end{equation}

The higher order cumulants are given by
\begin{equation}
	\kappa_{n\ge 2} = \left. \frac{\partial}{\partial t^n}
	\ln G_{\rm LO} (t)
	\right|_{t=0} = (n-2)!
\frac{ (N_c^2-1) S_\perp
\Lambda^2
 }{8 \pi} \left(\frac{ \mu_{\rm p}^2 {\mathfrak D}  }{\Lambda^{2}}\right)^{n}\,.
	\label{Eq:cn}
\end{equation}
Also note that owing to presence of the increasing powers of $1/\Lambda^2$,  $\kappa_{n+1} \gg \kappa_n$, and thus
the factorial cumulants defined by
\begin{equation}
	c_{n>2} = \left. \frac{\partial}{\partial z^n}
	\ln G_{\rm LO} (t=\ln z)
	\right|_{z=1}
\end{equation}
are approximately equal to the cumulants, $c_n \approx \kappa_n$.

\begin{figure}
\includegraphics[width=0.48\linewidth]{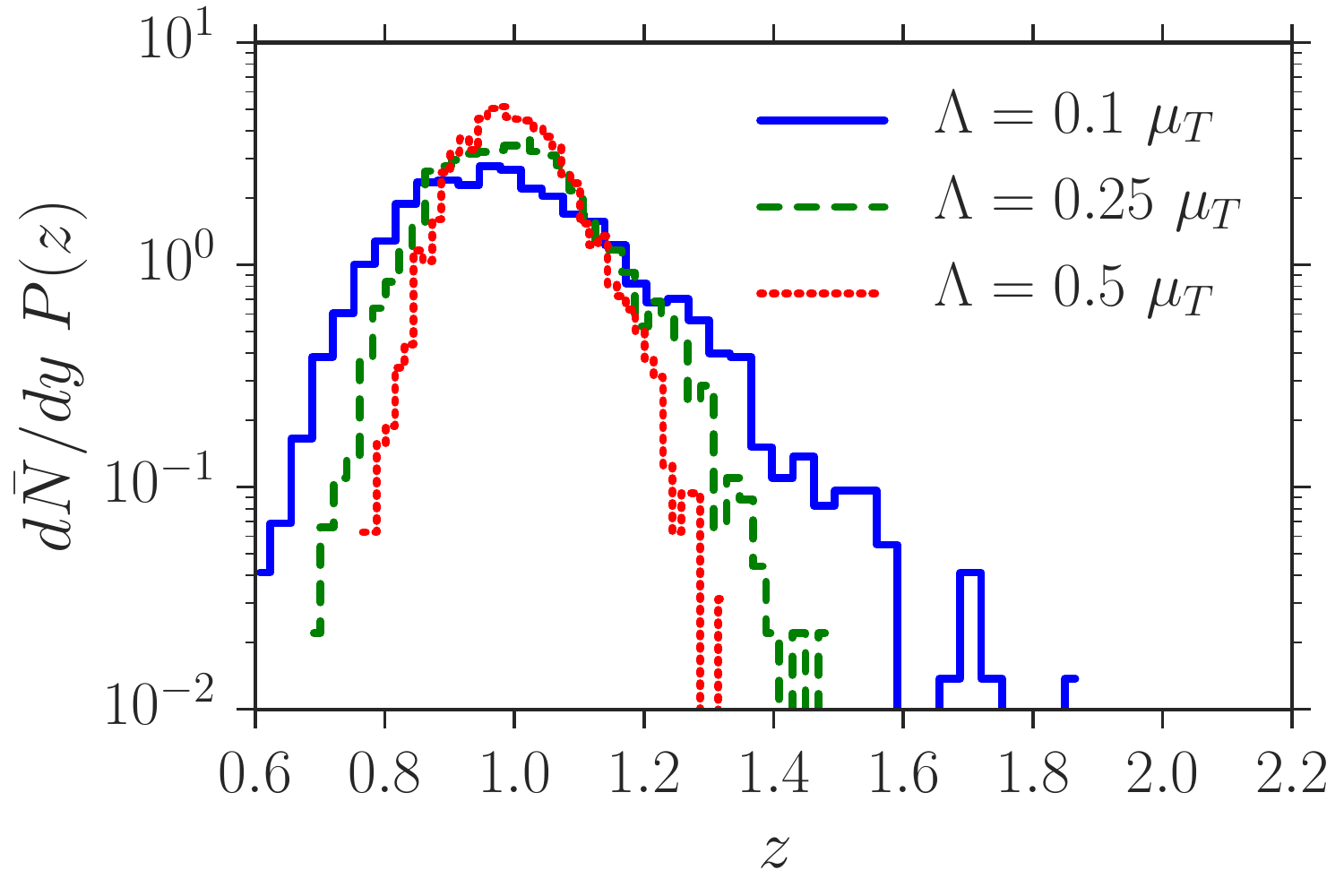}
\includegraphics[width=0.48\linewidth]{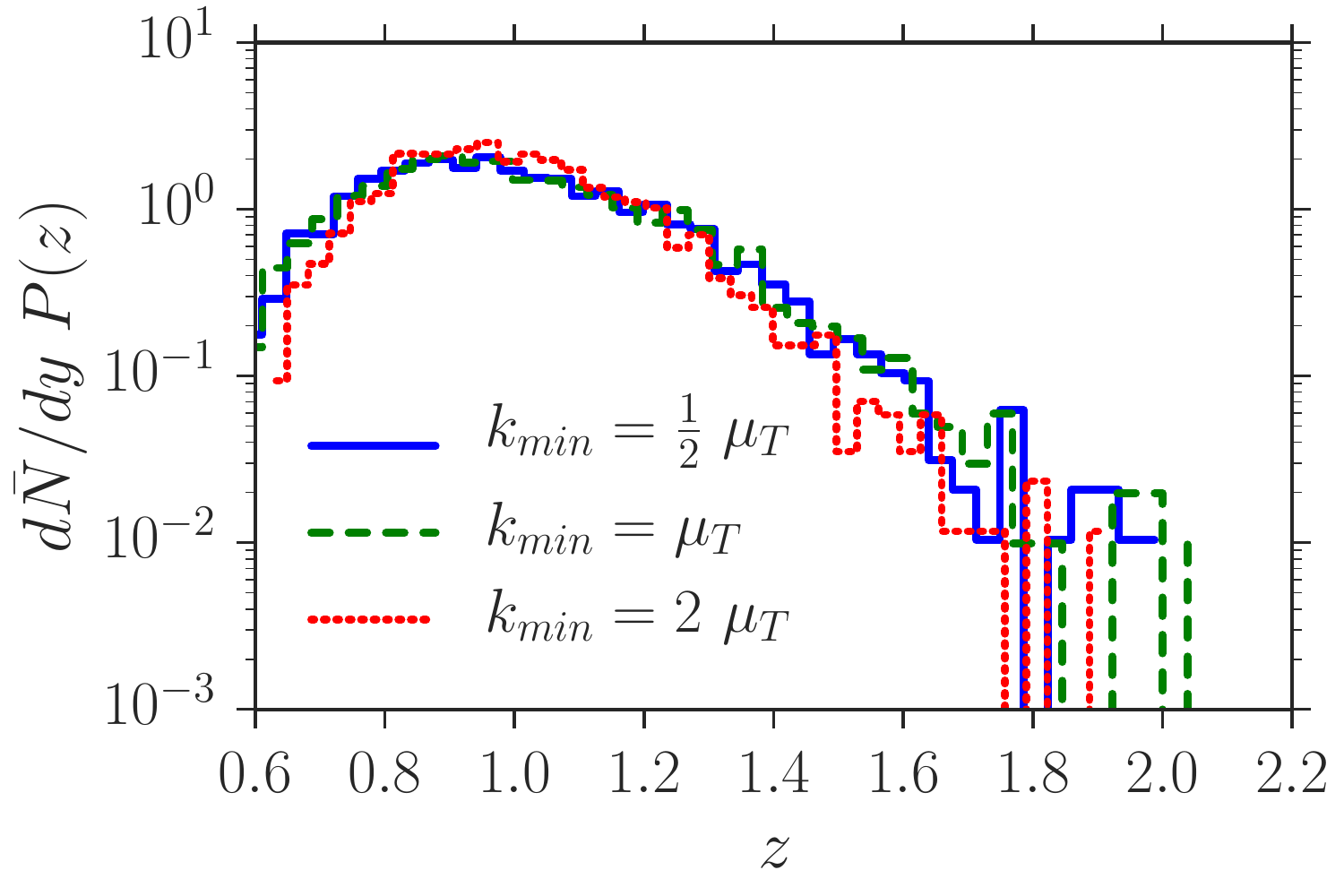}
\caption{ 
	The probability distribution multiplied by the average as a function of 
		$z=dN/dy / (d\bar{N}/dy)$   for different values of the infrared cut-off, $\Lambda$, and the minimal momentum of produced gluon,  $k_{\rm min}$.
	}
		\label{fig:KNO}
\end{figure}

The cumulants are very close to those of the $\gamma$-distribution
\begin{equation}\label{scaling}
	\bar x P(z=x/\bar x) = \frac{\alpha}{\Gamma(\alpha)}
	e^{-z\alpha} (\alpha z)^{\alpha-1}
\end{equation}
with
\begin{equation}
	\kappa_n = (n-1)!\ \alpha \left( \frac{\bar x}{\alpha} \right)^n\,.
\end{equation}
Since the $\gamma$-distribution is known to exhibit the KNO scaling,
we naturally expect to have KNO scaling at this order as well. We can check the KNO scaling by plotting the scaling function Eq.~(\ref{scaling}) for different values of $k_{\rm min}$ since varying  $k_{\rm min}$ one varies the mean multiplicity. This is plotted in Fig.~\ref{fig:KNO}, right panel. Note that what is plotted here is the  result of full numerical evaluation of the probability distribution (see Appendix for detail), and not just the leading approximation. The numerics is performed by using the MV model also for the target fields with the scale $\mu_{\rm t}\gg\mu_{\rm p}$. One observes that indeed for large enough $k_{\rm min}$ the quality of the  KNO scaling is very good.
 There is one subtlety here. Our distribution Eq.~(\ref{gd}) is not exactly the $\gamma$-distribution. In particular the first moment Eq.~(\ref{fm}) has an additional logarithmic factor relative to the parameter that determines the higher moments Eq.~(\ref{Eq:cn}). Since this logarithmic factor depends on $k_{\rm min}$, it  could potentially affect the KNO scaling. Nevertheless this additional logarithmic dependence is very slowly varying, and the scaling is clearly seen in the numerical results in
 Fig.~\ref{fig:KNO}.

 The left panel in Fig.~\ref{fig:KNO}  illustrates what happens when the soft scale $\Lambda$ is raised and becomes comparable with the target saturation momentum. One clearly observes that as $\Lambda$ grows, the probability distribution becomes narrower. Note that in this regime we cannot neglect the effects of HBT as well as correction due to exact form of the Lipatov vertex. These are the effects that drive the narrowing of the distribution for larger $\Lambda$.

The leading order probability distribution we obtained is similar in many respects to that obtained in  Ref.~\cite{Gelis:2009wh}. There are however some significant differences. First, in the case of dense target the contribution of the Bose enhancement in the target wave function is suppressed by the factor $1/N_c^2$, whereas in  Ref.~\cite{Gelis:2009wh}  the target was treated as dilute, this contribution was of order unity and contributed to the probability distribution on par with the projectile BE.

Second, the cumulants in our case are very close to those given by the gamma distribution, whereas  Ref.~\cite{Gelis:2009wh} finds negative binomial distribution (NBD). The  $n$-th cumulants of the two distributions differ by the factor $n-1$. This difference can be traced to a different way the function $\mu _{\rm p} ^2(p)$ is treated at small $p$ in the two approaches. Our calculation corresponds to taking constant $\mu _{\rm p} ^2$ and cutting off the putative infrared divergence in the integrals by a finite area of the projectile. The IR regulator therefore does not arise from making the correlation between the sources nonlocal in coordinate space, but rather imposing an impact parameter profile on the source density. On the other hand the authors of  Ref.~\cite{Gelis:2009wh} regulate IR divergences by taking $\mu _{\rm p} ^2(p)\sim p^2$ for momenta smaller than the projectile saturation momentum. Physically this corresponds to taking the correlation function between the two sources to be nonlocal in coordinate space.

This different treatment of the IR behavior leads to different integrals with respect to the incoming gluon from the
projectile wave function $q$. In this paper we keep $\mu_{\rm p}=$ const and thus
\begin{equation}
	\int d q^2 \frac{\mu_{\rm p}^{2n}} {q^{2n} } =  - \frac{1}{n-1} \frac{\mu_{\rm p}^{2n}}{ q^{2(n-1)} }\,,
	\label{Eq:qint}
\end{equation}
while the authors of Ref.~\cite{Gelis:2009wh} approximate $\mu_{\rm p}^2 \approx  q^2$; in this case
the integral in Eq.~\eqref{Eq:qint} brings no additional factors dependent on $n$.

Related to this point is also the different energy dependence that we expect from the distribution.
In our case the parameter of the distribution is $\alpha = \frac{N_c^2-1}{8\pi} S_\perp\Lambda^2$, where $S_\perp\Lambda^2\approx 1$. This parameter does not depend on
energy  for the dilute projectile. Thus we expect $\alpha$ to be approximately
energy independent. On the other hand the treatment of the IR in  Ref.~\cite{Gelis:2009wh} leads to replacement of $\Lambda^2$ by the saturation momentum of the projectile $Q_s^2$. This grows with energy, and thus the probability distribution has  rather strong dependence on energy.


\subsection{Constraint action formalism and the Liouville potential}
\label{sec:constr}

One interesting property of our derivation is that the probability distribution of produced particles that we find is deeply related with  the probability distribution of particles in the projectile wave function. Note that we have approximated the Lipatov vertex by its part that involves only the momentum of the gluon coming from the incoming wave function.  Thus if in the rest of the calculation we simply set $U(\v{x})=1$, that would correspond to calculating the particle distribution in the projectile  wave function.
This last move would simply correspond to setting $k_{\rm min}=0$ in Eq.~(\ref{funnyd}), since $\int \frac{d^2k}{(2\pi)^2}D(\v{k})=D(\v{x}=0)=1$ which is equivalent to setting $U=1$. This then gives
\begin{equation}\label{dm}
{\mathfrak D} =
\frac{2g^{2}}{\pi}\,.
\end{equation}
Clearly we should not set $k_{\rm min}=0$ in the integration limits in Eq.~(\ref{glo}), but instead keep it fixed. Thus we conclude that taking the limit of Eq.~(\ref{dm}) in Eq.~(\ref{glo}) corresponds to calculating the probability distribution for particles with transverse momenta smaller than momentum $k_{\rm min}$ in the projectile wave function.

Interestingly, such a calculation can be performed independently
using  an alternative formulation based on the
framework of the constraint effective potential, see Refs.~\cite{ORaifeartaigh:1986axd,KorthalsAltes:1993ca,Dumitru:2017cwt}.

The main idea of this approach is to
integrate out fluctuations of $\rho_{\rm p} (\v{q})$ that do not affect a specific
operator defined through  $\rho_{\rm p} (\v{q})$.
In Ref.~\cite{Dumitru:2017cwt} the constraint
effective potential for the gluon distribution defined by the covariant field, $A^+$,
was derived
\begin{equation}\label{eta}
	e^{-V_{\rm eff} [\eta(\v{q})] }
	= \frac{1}{Z_{\rm p}} \int {\cal D} \rho_{\rm p}
	W (\rho_{\rm p})
	\delta \left(  \eta(\v{q})   -
	\frac{
		g^2 {\rm tr} |A^+(\v{q})|^2}
		{  \langle g^2 {\rm tr} |A^+(\v{q})|^2 \rangle    } \right)\,,
\end{equation}
where $A^+(\v{q}) = g/q^2 \rho_{\rm p}(\v{q})$,
\begin{equation}
	\langle g^2 {\rm tr} |A^+(\v{q})|^2 \rangle
	= \frac12 (N_c^2-1) S_\perp \frac{g^4 \mu_{\rm p}^2}{q^4},
\end{equation}
and
\begin{equation}
	\label{Eq:Veff}
	V_{\rm eff} [\eta(\v{q})] = \frac{1}{2} (N_c^2-1) S_\perp \int \frac{d^2q}{(2\pi)^2}
	\left\{ \eta(\v{q}) -1 - \ln \eta (\v{q}) \right\}\,.
\end{equation}
This potential corresponds to a Liouville potential with negative Ricci scalar~\footnote{See also Ref.~\cite{Dumitru:2018iko} where this formalism was
applied to describe the centrality dependence of the nuclear modification factor.}.

The generating function for gluon multiplicity in the projectile is defined as
\begin{align}
	 G_{\rm LO} (t)
&=
  \left \langle\
 \exp\left[
 t
 \int_\Lambda^{k_{\rm min}}\frac{d^{2}q}{(2\pi)^{2}}
 \rho_{\rm p}^{a}(-\v{q})
\frac{
 {\mathfrak D} }{2 q^2}
\rho_{\rm p} ^{a}(\v{q})
 \right]\
\right\rangle_{\rm p} \,.
\end{align}
Using the effective potential \eqref{Eq:Veff} we have
\begin{align}
	\label{Eq:GLODapp}
	 G_{\rm LO} (t)
&=
  \left \langle\
 \exp\left[
 t
 \int_\Lambda^{k_{\rm min}}\frac{d^{2}q}{(2\pi)^{2}}
\rho_{\rm p} ^{a}(-\v{q})
\frac{
 {\mathfrak D} }{2 q^2}
\rho_{\rm p} ^{a}(\v{q})
 \right]\
\right\rangle_{\rm p}
\\\notag &
= \int {\cal D} \eta
\exp\left(-V_{\rm eff}[\eta(\v{q})]
+
\frac{1}{2} (N_c^2-1) S_\perp \int_\Lambda^{k_{\rm min}}\frac{d^{2}q}{(2\pi)^{2}}
\ t \frac{\mu^2_{\rm p}   {\mathfrak D}}{q^2} \
\eta(\v{q})
\right)
\\\notag &
=
\int {\cal D} \eta
\exp\left( -
\frac{1}{2} (N_c^2-1) S_\perp
\int \frac{d^2q}{(2\pi)^2}
	\left\{ \eta(\v{q}) -1 - \ln \eta (\v{q})	+ \theta(q-\Lambda) \theta(k_{\rm min} -q)
\ t \frac{\mu^2_{\rm p}   {\mathfrak D}}{q^2} \
\eta(\v{q})
 \right\}
\right) \,.
 \end{align}
For large $S_\perp$ this integral can be computed using the saddle point approximation
\begin{equation}
	\eta_{\rm SP}(\v{q}) = \begin{cases}
		\left( 1-   t \frac{\mu^2_{\rm p}   {\mathfrak D}}{q^2} \right)^{-1},\ {\rm if}\ \Lambda \le q \le k_{\rm min}\\
		1, \ {\rm otherwise}
	\end{cases}
\end{equation}
to yield
\begin{align}
	\label{Eq:GLODapp2}
	 G_{\rm LO} (t)
&=
 \exp\left[
 \frac{1}{2} (N_c^2-1) S_\perp
\int \frac{d^2q}{(2\pi)^2}
\ln  	\eta_{\rm SP}(\v{q})
\right]
\\\notag &
=
 \exp\left[
-	 \frac{1}{2} (N_c^2-1) S_\perp
	 \int_\Lambda^{k_{\rm min}} \frac{d^2q}{(2\pi)^2}
\ln
\left( 1-   t \frac{\mu^2_{\rm p}   {\mathfrak D}}{q^2} \right)
\right]
 \end{align}
which reproduces  the result obtained previously.
Interestingly, the origin of the logarithm in this equation is owing to the
presence of the Liouville logarithm in the effective constraint action for
$\eta(\v{q})$, see Eq.~\eqref{Eq:Veff}.

Note that the form of the integral appearing in Eq.~\eqref{Eq:GLODapp}  is quite suggestive,
\begin{equation}
	G_{\rm LO} (t) = 
\int {\cal D} \eta
	\exp\left(-V_{\rm eff}[\eta(\v{q})]
	+
t 	\frac{1}{2} (N_c^2-1) S_\perp \int_\Lambda^{k_{\rm min}}\frac{d^{2}q}{(2\pi)^{2}}
	\frac{\mu^2_{\rm p}   {\mathfrak D}}{q^2} \
	\eta(\v{q})
	\right) \,.
\end{equation}
Here $t$ can be viewed  as an external field, while
the moments and the cumulants of the gluon number can be viewed as
the moments and the cumulants for fluctuations of  the ``composite field''
$\int d^2 q \ \eta(\v{q})/q^2$. Thus the
fluctuations of the number of particles provide a direct measurement of the Liouville potential.

It is interesting to note that our derivation provides a concrete realization of early ideas in the literature about relevance of Liouville action to multiplicity fluctuations in CGC.
Ref.~\cite{Iancu:2007st} postulated {\it ad hoc} such a Liouville potential for {\it saturation momentum} fluctuations. In the present paper we instead derive it form the constrained
effective potential for the MV model~\footnote{Although our derivation was done in the MV model, the numerical result of Ref.~\cite{Dumitru:2017cwt}, see Fig. 5,
suggests that the high energy evolution does not change the form of the potential and only leads to the modification of the effective projectile area $S_\perp$.
This modification might be responsible for the origin of the effective width $\sigma$ used in Ref.~\cite{McLerran:2015qxa,Bzdak:2015eii}.
}.

 Although the argument of our potential is not the saturation momentum, but rather the composite filed $\eta$, they are closely related.
Consider the effective potential for  $\eta$ close to its saddle point
value at zero external field.
Expanding in $\ln \eta$ we obtain
\begin{equation}\label{lneta}
	V_{\rm eff}[\eta(\v{q})]  \approx
	\frac{1}{2} (N_c^2-1) S_\perp
	\int \frac{d^2 q}{(2\pi)^2} \frac{1}{2} \ln^2 \eta(\v{q})\,.
\end{equation}
Recall that the operator definition of the composite field $\eta$ is given by Eq.~(\ref{eta})
\begin{equation}
	\eta(\v{q})=\frac{\rho_{\rm p}^a(\v{q})\rho_{\rm p}^a(-\v{q})}{(N_c^2-1)\mu_{\rm p}^2(\v{q})}
\end{equation}
which is interpreted naturally as the scaled fluctuating saturation momentum of
the projectile, since on a configuration by configuration basis the saturation
momentum is indeed determined by the square of the color charge density. Thus we
indeed can interpret Eq.~(\ref{lneta}) as the probability distribution for the
fluctuating saturation momentum, which has exactly the same form as assumed in
Refs.~\cite{McLerran:2015qxa,Bzdak:2015eii}.



\section{Corrections to the generating function}
In this section we consider corrections to the cumulant generating function. There are two basic types of corrections: those attributed to BE with lower power
of IR divergence when integrating with respect to the incoming gluon from the
projectile wave function
and those attributed to the HBT contribution, see the section for two particle gluon production.
\subsection{Subleading BE terms}
The subleading IR terms due to Bose enhancement can be easily resummed.  We simply have to allow for the full Lipatov vertex and for the $q$-dependence of the dipole amplitude in Eq.~(\ref{glo}). Thus formally we get
\begin{eqnarray}\label{gnlo}
G_{\rm BE}(t)&=&\exp\left[-
	 \frac{1}{2}(N_c^2-1) S_\perp
	 \int_\Lambda^{k_{\rm min}}  \frac{d^{2}q}{(2\pi)^2}
\ln \left(
1 - t\  \mu_{\rm p}^2(\v{q}) \bar{\mathfrak D}(q)
\right)
 \right]\,,
\end{eqnarray}
where
\begin{equation}\label{funnierd}
	\bar{\mathfrak D}(q) =
\frac{4g^{2}}{(2\pi)^{3}}
\int_{k_{\rm min}} d^2 k D(\v{k}-\v{q})\Gamma(\v{k},\v{q},\v{q}) \,.
\end{equation}
The explicit form of the distribution now depends on the dipole amplitude $D(k)$, and can be calculated once this amplitude is known.

\subsection{The HBT contributions}
We now concentrate on the corrections of the second type. We will keep only
terms leading in $N_c$.

We rewrite our basic expression for the generating function as
\begin{align}
&	 G (t)
	=
\int \frac{{\cal D} \rho_{\rm p}}{Z_{\rm p}}
\int \frac{{\cal D} \rho_{\rm t} }{Z_{\rm t}}
	W_{\rm p}(\rho_{\rm p})
	W_{\rm t}(\rho_{\rm t})\\\notag& \times
\exp\Bigg[
 t
\frac{2g^{2}}{(2\pi)^{3}}\int\frac{d^{2}q}{(2\pi)^{2}}\frac{d^{2}q'}{(2\pi)^{2}}
\rho_{\rm p}^{a}(-\v{q}')
 \int d^2k\
\Gamma(\v{k},\v{q},\v{q}')
\left[U^{\dagger}(\v{k}-\v{q}')U(\v{k}-\v{q})\right]_{ab}
\ \rho_{\rm p}^{b}(\v{q})
 \Bigg] \\  & =
 \label{Eq:represented}
\int \frac{{\cal D} \rho_{\rm p}}{Z_{\rm p}}
		W_{\rm p}(\rho_{\rm p})
	\exp\Bigg[
 t
\frac{1}{2}\int\frac{d^{2}q}{(2\pi)^{2}}
\rho_{\rm p}^{a}(-\v{q})\bar{\mathfrak D}(q)
\ \rho_{\rm p}^{a}(\v{q})
 \Bigg]
	\\\notag& \times
	\int \frac{{\cal D} \rho_{\rm t} }{Z_{\rm t}}
	W_{\rm t}(\rho_{\rm t})
	\exp\Bigg[
 t
\frac{2g^{2}}{(2\pi)^{3}}\int\frac{d^{2}q}{(2\pi)^{2}}\frac{d^{2}q'}{(2\pi)^{2}}
\rho_{\rm p}^{a}(-\v{q}')
\Bigg\{ \int d^2k
\Gamma(\v{k},\v{q},\v{q}'):
\left[U^{\dagger}(\v{k}-\v{q}')U(\v{k}-\v{q})\right]_{ab}:
\Bigg\}
\rho_{\rm p}^{b}(\v{q})
 \Bigg] \,.
\end{align}
The BE  contribution is represented by the first line of Eq.~\eqref{Eq:represented}  is discussed
in the detail in previous section. The last two terms are the corrections we are after. We have introduced the normal ordered product of the Wilson lines
\begin{equation}
:\left[U^{\dagger}(\v{k}-\v{q}')U(\v{k}-\v{q})\right]_{ab}:\ \equiv
\left[U^{\dagger}(\v{k}-\v{q}')U(\v{k}-\v{q})\right]_{ab}
-
D(\v{k}-\v{q}) \delta_{ab} (2\pi)^2 \delta(\v{q}' - \v{q}).
\end{equation}

The normal ordering insures that apart from the BE term, no other terms contain contractions between two Wilson lines that belong to the same ``vertex''. These contractions have been completely resummed into the propagator of the color charge density
\begin{equation}
	\frac{1}{\hat{\mu}_{\rm p}^2(q)}
=
\frac{1}{{\mu}_{\rm p}^2}
-
 t\bar{\mathfrak D}(q)
\approx
\frac{1}{{\mu}_{\rm p}^2}
- t
\frac{
 {\mathfrak D} }{q^2}\,.
\end{equation}

 We can expand the functional integral into series in the ``interaction term''.
\begin{align}
&	 G (t)
	=
\int \frac{{\cal D} \rho_{\rm p}}{Z_{\rm p}}
	\exp\left[
-
\int\frac{d^{2}q}{(2\pi)^{2}}
\rho_{\rm p}^{a}(-\v{q})
\frac{1}{2 \hat{\mu}_{\rm p}^2}
\rho_{\rm p}^{a}(\v{q})
\right]
\int \frac{{\cal D} \rho_{\rm t} }{Z_{\rm t}}
W_{\rm t}(\rho_{\rm t})
	\exp\left[
 S_{\rm int} (\rho_{\rm p}, \rho_{\rm t})
\right]
\\ \notag & =
\int \frac{{\cal D} \rho_{\rm p}}{Z_{\rm p}}
	\exp\left[
-
\int\frac{d^{2}q}{(2\pi)^{2}}
\rho_{\rm p}^{a}(-\v{q})
\frac{1}{2 \hat{\mu}_{\rm p}^2}
\rho_{\rm p}^{a}(\v{q})
\right]
	\sum_n \frac{1}{n!}
	\int \frac{{\cal D} \rho_{\rm t} }{Z_{\rm t}}
W_{\rm t}(\rho_{\rm t})
 S^n_{\rm int} (\rho_{\rm p}, \rho_{\rm t}) \, ,
\end{align}
where the interaction part is given by
\begin{align}
	& S_{\rm int} (\rho_{\rm p}, \rho_{\rm t}) =
t
\frac{2g^{2}}{(2\pi)^{3}}\int\frac{d^{2}q}{(2\pi)^{2}}\frac{d^{2}q'}{(2\pi)^{2}}
\rho_{\rm p}^{a}(-\v{q}')
\Bigg\{ \int d^2k
\Gamma(\v{k},\v{q},\v{q}')
:\left[U^{\dagger}(\v{k}-\v{q}')U(\v{k}-\v{q})\right]_{ab}:
\Bigg\}
\rho_{\rm p}^{b}(\v{q})\,.
\end{align}

\begin{figure}
\includegraphics[width=0.18\linewidth]{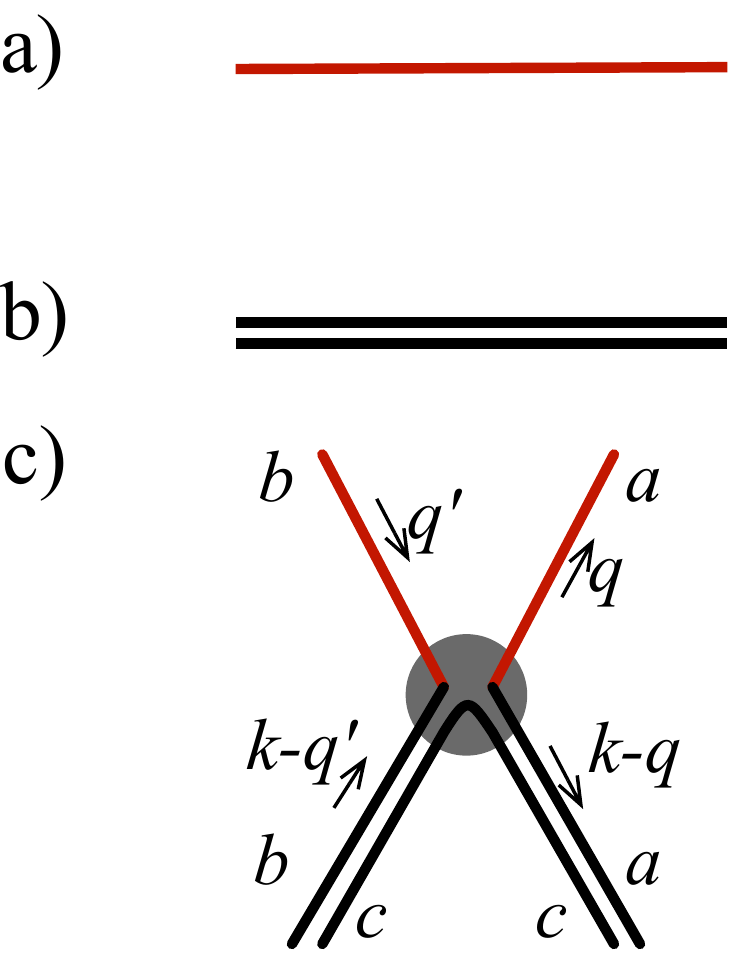}
\caption{Feynman rules for the propagators and the vertex;\\
	a) The resummed propagator of the projectile $\rho$: $\hat{\mu}^2_{\rm p}(\v{q})\delta_{ab}$; \\
	b) The propagator of the target  $U$: $D(\v{q})/(N_c^2-1)\delta_{bc}\delta_{cd}$; \\
	c) The ``interaction  vertex'':    $t \frac{8 g^2} { (2\pi)^3} \Gamma(\v{k}, \v{q},\v{q}')
  $.
}
\label{fig:FR}
\end{figure}

To compute the cumulant generating function it is  useful to introduce the Feynman rules depicted in Fig.~\ref{fig:FR}. Note that due to normal ordering in the vertex, no diagram with contraction of the two $U$'s belonging to the same vertex are allowed. Those have already been resummed in the propagator of $\rho _{\rm p}$.

\begin{figure}
\includegraphics[width=0.48\linewidth]{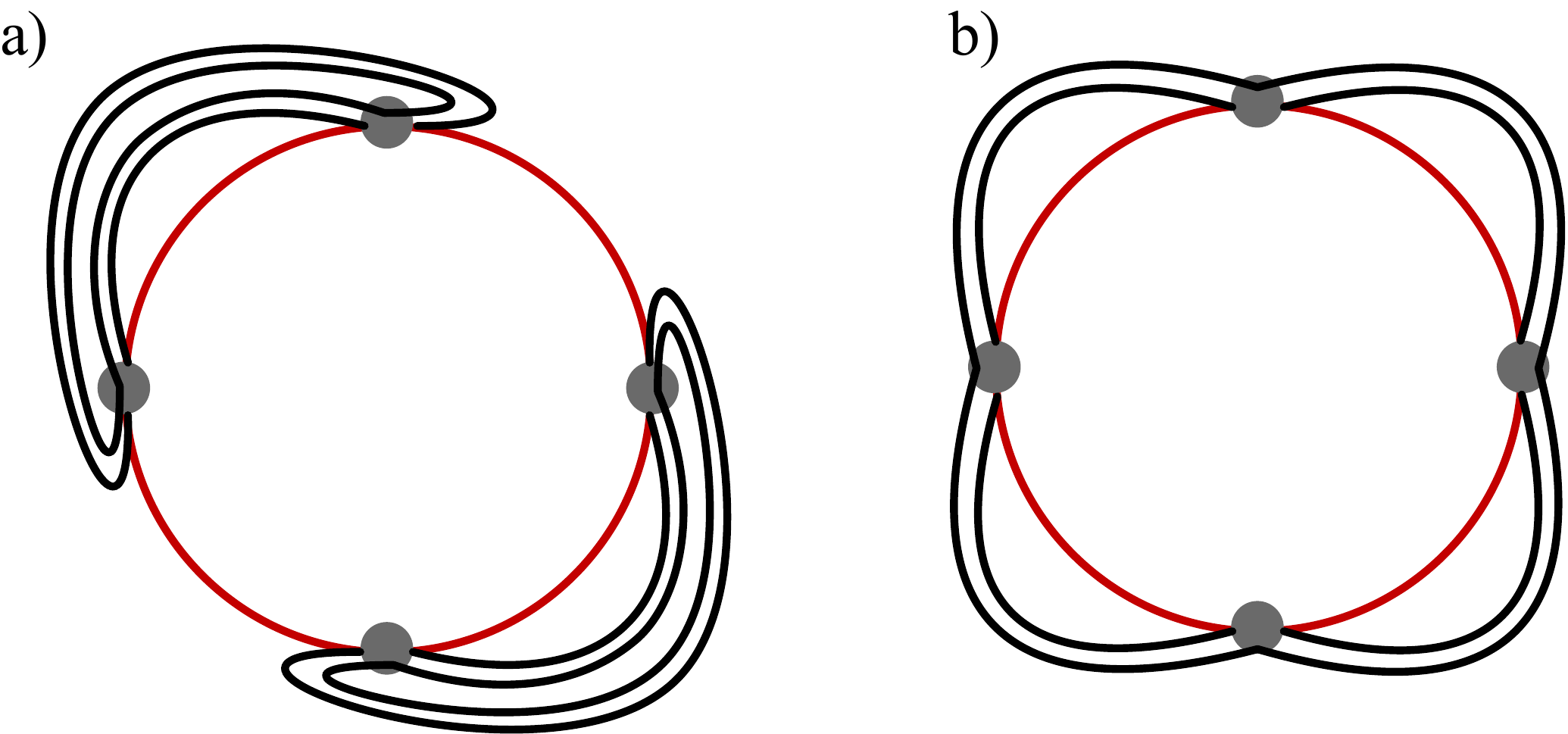}
\caption{Example of the leading $N_c$ contribution.}
\label{fig:Romashka}
\end{figure}

To understand which type of diagrams give the leading corrections, let us first consider a particular example: the connected contribution involving four vertexes (correction to inclusive four particle production).
The leading $N_c$ contributions have two different topologies, see
Fig.~\ref{fig:Romashka}.

The diagram a) in  Fig.~\ref{fig:Romashka} is given by (modulo combinatorial and kinematic factors)
\begin{align}
	\int d^2k_1
     d^2k_2
	\int \frac{d^2 q}{(2\pi)^2}
	 \frac{d^2 q_1}{(2\pi)^2}
	 \frac{d^2 q_2}{(2\pi)^2} &
	\Gamma^2(\v{k}_1, \v{q},\v{q}_1)
	\Gamma^2(\v{k}_2, \v{q},\v{q}_2)
	\hat{\mu}^4_{\rm p}(\v{q})
	\hat{\mu}^2_{\rm p}(\v{q}_1)
	\hat{\mu}^2_{\rm p}(\v{q}_2)\\ \notag &\times
	D(\v{k}_1-\v{q})
	D(\v{k}_1-\v{q}_1)
	D(\v{k}_2-\v{q})
	D(\v{k}_2-\v{q}_2)\,.
\end{align}
The diagram b) in  Fig.~\ref{fig:Romashka} is given by
\begin{align}
	\int d^2k
	\int \frac{d^2 q_1}{(2\pi)^2}
	 \frac{d^2 q_2}{(2\pi)^2}
	\frac{d^2 q_3}{(2\pi)^2}
	\frac{d^2 q_4}{(2\pi)^2}
	&\Gamma(\v{k}, \v{q}_1,\v{q}_2)
	\Gamma(\v{k}, \v{q}_2,\v{q}_3)
	\Gamma(\v{k}, \v{q}_3,\v{q}_4)
	\Gamma(\v{k}, \v{q}_4,\v{q}_1)
	\hat{\mu}^2_{\rm p}(\v{q}_1)
	\hat{\mu}^2_{\rm p}(\v{q}_2)
	\hat{\mu}^2_{\rm p}(\v{q}_3)
	\hat{\mu}^2_{\rm p}(\v{q}_4)
	\\ \notag &\times
	D(\v{k}-\v{q}_1)
	D(\v{k}-\v{q}_2)
	D(\v{k}-\v{q}_3)
	D(\v{k}-\v{q}_4)\,.
\end{align}
As before we  consider the IR dominant contribution from the Lipatov vertices.
For the diagram b) we get
\begin{equation}
 	\Gamma(\v{k}, \v{q}_1,\v{q}_2)
	\Gamma(\v{k}, \v{q}_2,\v{q}_3)
	\Gamma(\v{k}, \v{q}_3,\v{q}_4)
	\Gamma(\v{k}, \v{q}_4,\v{q}_1)
	\approx
	\frac{ \v{q}_1 \cdot \v{q}_2 }{q_1^2 q_2^2}
	\frac{ \v{q}_2 \cdot \v{q}_3 }{q_2^2 q_3^2}
	\frac{ \v{q}_3 \cdot \v{q}_4 }{q_3^2 q_4^2}
	\frac{ \v{q}_4 \cdot \v{q}_1 }{q_4^2 q_1^2}
\end{equation}
which after angular integration has logarithmic divergence for each integral of the form  $\int d q_i^2 /q_i^2$.
For the diagram a) we get
\begin{equation}
 	\Gamma^2(\v{k}_1, \v{q},\v{q}_1)
	\Gamma^2(\v{k}_2, \v{q},\v{q}_2)
	\approx
	\left(\frac{ \v{q} \cdot \v{q}_1 }{q^2 q_1^2}
	\frac{ \v{q} \cdot \v{q}_2 }{q^2 q_2^2}\right)^2
\end{equation}
and in this case the integral with respect to $q$ has quadratic divergence in IR. This divergence is of course regulated by $\Lambda^2\propto 1/S_\perp$
which leads to the extra power of $S_\perp$.

 It is now clear what are the leading diagrams due to the HBT corrections that contribute most to the generating function. Those are the diagrams that contain maximal number of the $\rho_{\rm p} $ propagators at the same momentum $q$, since each such propagator is accompanied by a product of two Lipatov vertexes containing the same momentum $q$, thereby leading to one extra power of area for each additional $\rho _{\rm p} $ propagator. These diagrams are of the type of Fig.~\ref{fig:Romashka} a, where every the vertexes are organized into pairs with two vertexes of the pair  connected by two propagators of the $U$ field, and one propagator of $\rho _{\rm p} $. Physically this corresponds to contributions to the $n$-gluon inclusive production, where gluon pairs are emitted independently, but the HBT correlations are present between two gluons in each pair.

 We are now going to ignore diagrams of type B but resum all diagrams of type A, corresponding to pairwise HBT correlations.

 For a diagram with $2n$ vertexes we have the following combinatorial coefficient
 $\frac{1}{(2n)!}$ from expanding the exponential;
  $(2n-1)!!$ - the number of ways to organize the $2n$ vertexes into pairs, which then will be contracted over $U$;
 $2^n$ - the number of contractions of $U$ - two possibilities within each pair of vertexes;
  $2^n$ - the number of contraction between two $\rho_{\rm p} $'s within the pair of vertexes. Although there are 4 possible of contractions in general, once a particular contraction of $U$'s is chosen, only two contractions of $\rho_{\rm p} $'s are leading order in $N_c$;
  $\frac{(n-1)!}{2}$ ways of ordering the $n$ vertex pairs along a circle. This is not just a number of permutations $n!$, since the position of the first vertex should be fixed (periodicity along the circle)-  that given $(n-1)!$, and the factor $1/2$ is due to identical nature of a permutation and its reflection, since in both cases every vertex has identical neighbors, and that's all that matters;
  $2^n$ way to contract the factors of $\rho _{\rm p}  $ along the circle, since for each vertex each one of two $\rho _{\rm p} $'s can either be contracted with its right or left neighbor;
 the color factors all cancel out except for the overall $N_c^2-1$ in front for any $n$.

Thus at the end of the day the diagram with $2n$ vertexes which are contracted pairwise into the ``daisy'' is $2^{2n-1}\frac{1}{n}$.

Resumming these terms we obtain
\begin{align}
	\ln G(t) - \ln G_{\rm LO} (t) &\approx
	\frac{(N_c^2-1)}{2} S_\perp
	\int_\Lambda^{k_{\rm min}} \frac{d^2 q}{(2\pi)^2}\
	\sum_{n=1}^\infty \frac{1}{n}
	{\cal Z}^{n}(\v{q}) \hat\mu^{2n}_{\rm p}(\v{q})
	\\\notag & =-
	\frac{(N_c^2-1) S_\perp}{2}
	\int_\Lambda^{k_{\rm min}} \frac{d^2 q}{(2\pi)^2} \ln\left[1-
		t^2 \frac{
		\hat\mu^{2}_{\rm p}(\v{q})
	{\cal Z}  }{q^2} \right ]
	\label{Eq:Hrenzantema}
\end{align} with
\begin{equation}
	\\ {\cal Z}
 =
	\int_\Lambda^{k_{\rm min}} \frac{d^2 q'}{(2\pi)^2} \mathfrak{D}_2(q')
	\frac{\hat{\mu}^2_{\rm p} (\v{q}')}{{q'}^2}
\end{equation}
and
\begin{equation}
 \mathfrak{D}_2(q)\equiv
 \left(
 \frac{ g^2}{\pi^3}
 \right)^2
 \int 
 {
 d^2k
 }
 D(k)D(k-q)\,.
\end{equation}

Finally we obtain the cumulant generating function in the form
\begin{align}
& \ln G(t)=-
	\frac{(N_c^2-1) S_\perp}{2}
	\int_\Lambda^{k_{\rm min}} \frac{d^2 q}{(2\pi)^2} \ln\left\{\mu^2_{\rm p}(\v{q})\left[\frac{1}{\hat\mu^{2}_{\rm p}(\v{q})}-
	t^2 \frac { {\cal Z} } { q^2}\right ]\right\}\,
	\\ & \notag
	=
\frac{(N_c^2-1) S_\perp}{2}
	\int_\Lambda^{k_{\rm min}} \frac{d^2 q}{(2\pi)^2} \ln\left\{
		1 - t
		\frac{\mu^2_{\rm p}}{q^2} \left( {\mathfrak D} + t Z\right)
		\right\}\,,
\end{align}
which has the same form as Eq.~\eqref{gd};
thus the momentum integral can be performed analytically leading to Eq.~\eqref{agd} with the substitution
 $ {\mathfrak D} \to   \left( {\mathfrak D} + t Z\right)$.

This is a rather simple expression, and one can analyze the effects of the correction given a model for the dipole amplitude $D(p)$.
These corrections may be important in the regime where $k_{\rm min}$ is not significantly greater than the soft scale $\Lambda$. Although we do not consider such a situation in the present paper, at high enough energy the projectile wave function itself will acquire a saturation momentum scale $Q_{\rm s, p}$ significantly larger than $\Lambda$. In this case it is quite conceivable that in our expressions the soft scale $\Lambda$ will be replaced by this semi soft scale $Q_{\rm s, p}$. In this case it is perfectly sensible to consider $k_{\rm min}<Q_{\rm s, p}$. The HBT contributions in this regime will become significant, and the relative significance of the BE and HBT contributions to the multiplicity fluctuations has to be reanalyzed.
We will not attempt to do it in the present paper.

\section{Conclusions}

In this paper we studied the multiplicity fluctuations in p-A collisions within the framework of dense-dilute CGC formalism using the MV model for the wave function of the proton.
Our approach is similar in many aspects to that of Ref.~\cite{Gelis:2009wh}. There are however some significant differences, and our results are quite different as well. As opposed to Ref.~\cite{Gelis:2009wh} we treat the target as very dense, while on the projectile side we do not assume any dynamical ``correlation length'' associated with the saturation momentum. We rather treat the IR physics of the proton wave  function as genuinely non perturbative, governed by a soft scale of order of the inverse proton size. As a result we obtain the probability distribution which within large range of energies is energy independent.

We identified two sources of multiplicity fluctuations: those due to the Bose enhancement of gluons  in the proton wave function and the HBT effect in the initial stages of scattering. Interestingly, in the dense-dilute framework the Bose enhancement in the nucleus wave function leads only to a $(N_c^2-1)^{-1}$ suppressed contribution to any cumulant of particle number, and is thus a subleading effect.
We demonstrated that as long as the low momentum structure of the proton wave function is dominated by the genuine soft scale (the ``proton size''), the dominant origin for the multiplicity fluctuations is the Bose enhancement.

We have calculated explicitly the moment generating function for the multiplicity distribution due to BE. The distribution we obtain is very close to the $\gamma$-distribution. Just like the $\gamma$-distribution it satisfies the KNO scaling with very good precision.
Interestingly, the leading term in the generating function for multiplicity of produced particles is practically identical to the generating function for multiplicity distribution in the projectile wave function. This latter quantity can be calculated using the effective action approach as suggested in Ref.~\cite{Dumitru:2017cwt}. We have shown that this effective action is nothing but the Liouville action for the composite field, which can be thought of as fluctuating density (or saturation momentum).

The authors of Ref.~\cite{Gelis:2009wh} obtained the negative binomial distribution for the multiplicity. Our result, as mentioned above is somewhat different, although for large moments the NBD and the $\gamma$-distribution are quite similar. The main difference, as explained in the text is in the physics of the scale that regulates the formal IR divergences.  In the dense - dilute calculation performed in the present paper this role is played by the soft scale of the proton radius, and not by the semi soft scale of the projectile saturation momentum.
If one assumes that the proton wave function itself is characterized by a finite correlation length, much smaller than the proton size one would have to reanalyze to what extent the dominance of the BE persists. It may well happen that the HBT contributions become equally important and have to be included in the leading order calculation. In fact this is precisely what happens on the target side, where the presence of large saturation momentum strongly suppresses the BE effect, as we noted above.
 It is thus not clear to us that the approximation of BE dominance and finite saturation momentum of the projectile are mutually compatible.

We note that both NBD, and $\gamma$-distribution have rather long tails for large values of produced multiplicity. It is very natural that these tails are associated with the quantum Bose enhancement effect of identical gluons, just like the Bose-Einstein distribution of identical noninteracting bosons. Thus we believe that although the details of the distribution are model dependent (MV model in our case), the main feature of large fluctuations is universal as long as the fluctuations are dominated by BE.

Finally, we have also calculated the correction due to the generating function due to pairwise HBT correlations. In the regime studied in the present paper this correction is small. However it is bound to become important in the regime of saturated projectile, and therefore in itself would be an interesting object of study.

\appendix

\section{Numerics}
For numerical calculations it is easier to work in the following semi-factorizable
representation (see Ref.~\cite{McLerran:2016snu})
\begin{align}
\left.\frac{dN}{d^{2}qdy}\right|_{\rho_{\rm p},\rho_{\rm t}} &
=\frac{2}{(2\pi)^{3}}\frac{1}{|\v{q}|^{2}}\left(\delta_{ij}\delta_{lm}+\epsilon_{ij}\epsilon_{lm}\right)\Omega_{ij}^{a}(\v{q})\left[\Omega_{lm}^{a}(\v{q})\right]^{*}\nonumber \\
 & =\frac{2}{2(2\pi)^{3}}\frac{1}{|\v{q}|^{2}}\left(\Omega_{\|}^{a}(\v{q})\left[\Omega_{\|}^{a}(\v{q})\right]^{*}+\Omega_{\perp}^{a}(\v{q})\left[\Omega_{\perp}^{a}(\v{q})\right]^{*}\right)\label{Eq:LO}
\end{align}
where
\begin{equation}
	\Omega_{ij}^{a}(\v{q})=
	\int d^2 x e^{-i\v{q}\cdot\v{x}}\,
	\Omega_{ij}^{a}(\v{x})
\end{equation}
and
\begin{equation}
	\Omega_{ij}^{a}(\v{x})=g\left[\frac{\partial_{i}}{\partial^{2}}\rho_{\rm p}^{b}(\v{x})\right]\partial_{j}U^{ab}(\v{x})\label{Eq:Omega}
\end{equation}
with the adjoint Wilson line defined as
\begin{equation}
U^{ab}(\v{x})=2{\rm tr}\left[t^{b}V^{\dagger}(\v{x})t^{a}V(\v{x})\right]\label{Eq:Adj}
\end{equation}
and
\begin{align}
\Omega_{\|}^{a}(\v{k}) & =\delta_{lm}\Omega_{lm}^{a}(\v{k})\equiv\Omega_{11}^{a}(\v{k})+\Omega_{22}^{a}(\v{k})\,,\\
\Omega_{\perp}^{a}(\v{k}) & =\epsilon_{lm}\Omega_{lm}^{a}(\v{k})\equiv\Omega_{12}^{a}(\v{k})-\Omega_{21}^{a}(\v{k})\,.
\end{align}

The calculations are performed on a two dimensional lattice.
For the projectile the color sources are generated from a
Gaussian ensemble, see Eq.~\eqref{Eq:MVp}, with $\mu_{\rm p}/\mu_{\rm t} = 1/4$
and the radius of the projectile, $R_{\rm p}  = 1/\mu_{\rm p}$.
The Poisson equation entering in Eq.~\eqref{Eq:Omega},
$\frac{1}{\partial^{2}}\rho_{\rm p}^{a}(\v{x})$
is regulated by $\Lambda$, according to
\begin{equation}
	\frac{1}{\partial^{2}}\rho_{\rm p}^{a}(\v{x}) \to
	\frac{1}{\partial^{2}-\Lambda^2}\rho_{\rm p}^{a}(\v{x})\,.
\end{equation}

For the target, we  use the MV model with
\begin{equation}
	\langle \rho_{\rm t}^a(x^-, \v{x}) \rho_{\rm t}^b(y^-, \v{y})  \rangle_{\rm t}
	=   \mu_{\rm t} ^2   \delta(\v{x}-\v{y}) \delta(x^--y^-) \delta^{ab}
	\label{Eq:MVt}
\end{equation}
and compute the fundamental Wilson lines
\begin{equation}
	V(\v{x}) =
	\mathbb{P}
	\exp \left(
	i g^2 \int dx^-  t^a  \frac{1}{\partial^2}\rho _{\rm t} ^a (x^-,\v{x})	\right) \, .
	\label{Eq:Wilson}
\end{equation}
Further details can be found in
Refs.~\cite{Lappi:2007ku,Dumitru:2014vka}.

\section*{Acknowledgement}
We thank
Anton Andronic,
Adam Bzdak,
Michael Lublinsky,
Mark Mace,
Larry McLerran,
Prithwish Tribedy
and Urs Wiedemann
for  useful discussions on problems related to this project.
We are especially thankful to Adrian Dumitru, Yuri Kovchegov
and R. Venugopalan for illuminating discussions.

V.S. thanks Peter Braun-Munzinger for questions and discussions
during the EMMI Nuclear Matter and Particle seminar. This paper is
an attempt to answer one of Peter's question.

This research was supported by the NSF Nuclear Theory grant 1614640  and CERN scientific associateship (A.K.).
V.S. is indebted to Urs Wiedemann and CERN TH group for the partial support and hospitality at CERN, where this work was finalized.
V.S. also gratefully acknowledges partial support by the ExtreMe Matter Institute EMMI
(GSI Helmholtzzentrum f\"ur Schwerionenforschung, Darmstadt, Germany).

\bibliography{v3}
\end{document}